\input{aipcheck.tex}

\def\selectedoptions{final}
\documentclass[
   \selectedoptions
  ]
  {aipproc}

\def\selectedlayoutstyle{6x9}
\layoutstyle\selectedlayoutstyle

\SetInternalRegister\hbadness{8000} % pseudo latin isn't breaking very well :-)

\newcommand\doingARLO[2][]{%
  \ifx\mmref\undefined #1\else #2\fi
}

\begin{document}

\title 
      [$W$ Charge Asymmetry Measurement in CDF Run\,2]
      {$W$ Charge Asymmetry Measurement in CDF Run\,2}

\classification{43.35.Ei, 78.60.Mq}
\keywords{Document processing, Class file writing, \LaTeXe{}}

\author{C. Issever for the CDF collaboration}{
  address={University of California, Physics Department, Santa Barbara, CA 93106},
  email={issever@fnal.gov},
  thanks={This work was commissioned by the AIP}
}

%\iftrue
%\author{Arno Mittelbach}{
%  address={Zedernweg 62, 55128 Mainz, Germany},
%  email={arno@mittelbach-online.de},
%}

%\author{D. P. Carlisle}{
%  address={Willow House, Souldern},
%  email={david@dcarlisle.demon.co.uk},
%  homepage={http://www.dcarlisle.demon.co.uk},
%  altaddress={When I go to work: NAG Ltd, Oxford}
%}
%\fi

% \copyrightholder{Acoustical Scociety of America}
\copyrightyear  {2002}

\begin{abstract}
We present the status of the forward-backward charge
asymmetry measurement for $W$ boson production using early Run\,2 data
collected with the Collider Detector at Fermilab (CDF).
Tracking for forward electrons is a critical component of this
measurement, and we describe a new technique which combines the position
and energy measurements from the calorimeter with position measurements
in the silicon detector to provide tracking and charge determination for
electron candidates.
The performance of this algorithm is described and the sensitivity
for the $W$ charge asymmetry measurement with Run\,2 data is quantified.
\end{abstract}

\date{\today}

\maketitle

\section{Introduction}

Measurement of the forward-backward charge asymmetry in
$p \bar p \rightarrow W^{\pm}$ provides a constraint on the parton fluxes 
within the proton.
Since $u$ quarks carry, on average, a higher fraction of the proton
momentum than $d$ quarks, the $W^+$ in $u \bar d \rightarrow W^+$ tends to
be boosted in the proton direction.
Similarly, a $W^-$ tends to be boosted in the anti-proton direction.
This results in an expected non-zero forward-backward
charge asymmetry defined to be
\begin{equation} \label{eqwasymmetry}
A(y_W)=\frac{d\sigma(W^+) / dy - d\sigma(W^-) / dy}
{d\sigma(W^+) / dy +d\sigma(W^-) / dy}\,,
\end{equation}
where $y$ is the rapidity of the $W$ bosons and 
$d\sigma(W^{+,-})/dy$ is the differential cross section for
$W^+$ or $W^-$ boson production\footnote{The rapidity is defined by
$y=\frac{1}{2}\ln{(\frac{E+p_z}{E-p_z})}$ with the four-momentum of the particle
$p=(E,p_x,p_y,p_z)$.}. 

\begin{figure}
\includegraphics[width=0.6\textheight]{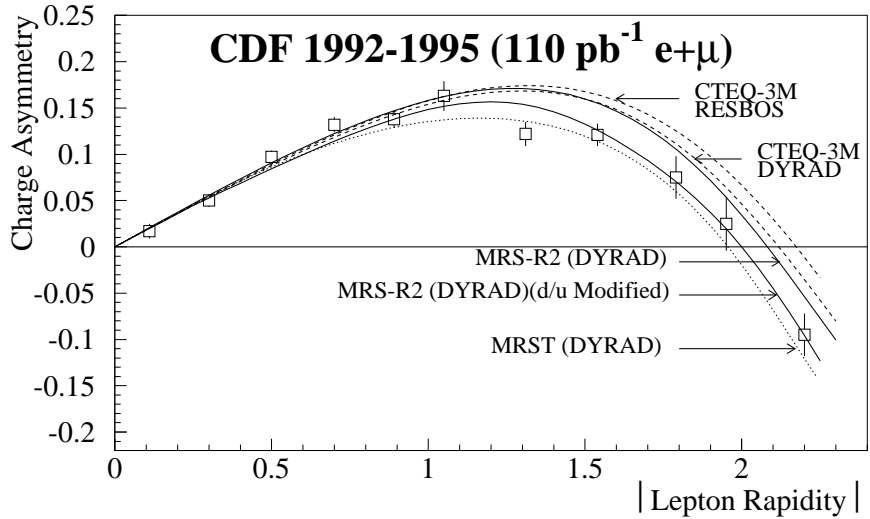}
\caption{CDF's lepton charge asymmetry measurement from Run\,1 is compared to
the predictions of various parton distribution functions as a function of lepton
rapidity. The discrimination is strongest at
large values of $|{\rm Lepton Rapidity}|$\cite{Abe:1998}. \label{figrun1-wa}}
\end{figure}

Leptonic decays of the $W$ boson provide a cleanly
identified sample for studying this asymmetry. Here we consider the
electron mode, $W^\pm \rightarrow e^\pm \nu$.
Because the neutrino escapes detection, the rapidity of the $W$ bosons is
not directly measurable and the $e^\pm$ direction is used instead.
Furthermore, the pseudo-rapidity $\eta$ is used to provide a simple but
good approximation of the rapidity\footnote{
The coordinate system is such that the polar angle $\theta$ is
measured from the proton direction, the azimuthal angle $\phi$ is
measured from the Tevatron plane and the pseudo-rapidity is defined
as $\eta=-ln(\tan{(\theta/2)})$.}.
The forward-backward lepton asymmetry,
\begin{equation} \label{eqleptonasymmetry}
A(\eta_l)=\frac{d\sigma(e^+) / d\eta - d\sigma(e^-) /
d\eta}{d\sigma(e^+) / d\eta +d\sigma(e^-) / d\eta}\,, 
\end{equation}
provides an experimental observable which convolves the $W$ production
asymmetry with the V-A decay distribution.
$A(\eta_l)$ (and also $A(y_W)$) is sensitive to the ratio of the
parton density functions for $u$ and $d$ quarks, $u(x)/d(x)$.
This sensitivity is most pronounced at high values of $|\eta|$
(forward region) as shown in Fig. \ref{figrun1-wa}.

Thus the primary experimental challenge in this measurement is tracking
in the forward region to obtain $e^\pm$ charge identification.
We describe below a newly developed algorithm which uses CDF's
extended silicon tracking coverage to reconstruct electron trajectories
in the forward direction.
This approach is applicable to many electron based measurements in addition to
this first use for the $W$ charge asymmetry measurement.
We quantify the sensitivity for the asymmetry measurement with the early data
from Run\,2 and extrapolate the sensitivity to the larger data samples which will
be collected soon.

\section{The CDF Run\,2 Detector} \label{secdetector}
The Collider Detector at Fermilab underwent a major upgrade
program for Run\,2 which is described in detail elsewhere\cite{detector:1996}. 
The features which are particularly relevant to this analysis are
a new, scintillating tile based, end-plug calorimeter, a multi-wire drift
chamber (COT), and a substantially extended silicon tracking system.

Electrons are identified by energy deposition in the calorimeters
which measure energy in separate
electromagnetic (EM) and hadronic (HAD) sections.
Position sensitive detectors (PES) located in the EM section 
at the expected maximum shower depth measure the position of electron
candidates with a precision of about 1\,mm.
Calorimeter based selection requirements are applied to electron candidates
to reject jet backgrounds, but to significantly reduce the backgrounds track
matching is needed.

\begin{figure}
\includegraphics[width=0.5\textheight]{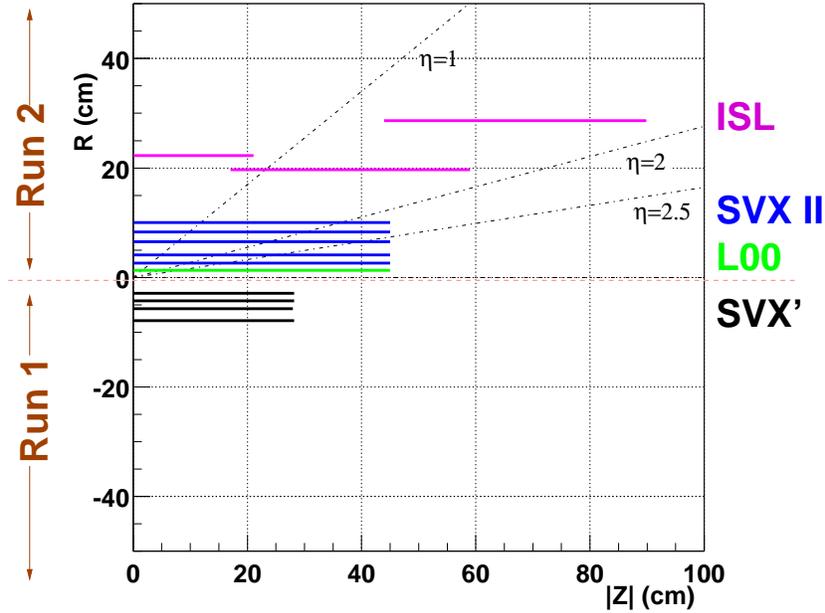}
\caption{Comparison of the Run\,2 and Run\,1 silicon detector in a
schematic rz-view. Lower part: The Run\,1 Silicon Vertex Detector
($\rm{SVX}^{\prime}$) with 96 single-sided ladders. Upper part:Run\,2 silicon detector consisting out of
three subdetectors: Layer 00 (L00) a single-sided layer of 48 ladders mounted directly on
the beampipe, which enhances the impact parameter resolution. The
Silicon Vertex Detector (SVX\,II)
with 360 double-sided ladders and the Intermediate Silicon Layers (ISL) with 296
double-sided ladders are providing precise
hits over a large lever arm out to $|\eta| \simeq 2$ and three dimensional tracking. \label{figsiliconrz}}
\end{figure}

The momentum and charge of final state particles are precisely measured by
track reconstruction in the COT, but its acceptance is limited to $|\eta|<1$.
Track reconstruction in the higher $|\eta|$ region 
is provided by the silicon system.
It consists of 8 measurement
layers spanning a radial region from 1.3\,cm to 28\,cm.
As shown in Fig.~\ref{figsiliconrz}, the coverage is substantially increased
relative to Run\,1. The large radial lever arm and length provide precision
tracking out to $|\eta|\sim 2$. This nearly doubles the tracking coverage
from the COT region and provides charge identification in the region which is
important for this measurement.

\section{Calorimeter Seeded Silicon Tracking} \label{secphoenix}

To fully exploit the forward coverage of the silicon tracking system,
we have developed a calorimeter-seeded silicon tracking algorithm which
reconstructs the trajectories of electrons without the COT.
In the central region, tracks reconstructed with the COT are improved
by the addition of silicon hit information.
The COT tracks are projected into the silicon where hits in a narrow road around
the projection are considered for addition to the fit.
Similarly, the new algorithm uses information from the calorimeter to seed the silicon
track reconstruction.
It is also possible to reconstruct tracks using only the silicon information.
That capability is important for tagging heavy flavor decays in the forward
direction, but seeding the fit with calorimeter or COT information is
more robust.

Electron candidates are first identified based on calorimeter measurements.
The energy deposited in the hadron section of the calorimeter
is required to be less than $5\%$ of the energy in the electromagnetic
section, and the extra energy in a cone of radius 0.4 around the primary deposition point
is required to be less than $10\%$ of the total.
The centroid of the energy clustered in the PES is used to determine the
position of the electron candidate in the calorimeter.
The position of the primary collision vertex is measured with the
other tracks in the event and provides a second point for the electron trajectory.
The curvature of the electron trajectory is determined from the transverse energy,
$E_t = E \sin{\theta}$, measured by the calorimeter.
These two points and the curvature, with appropriate covariance,
are used to generate a seed helix for each charge hypothesis.

\begin{figure}
\includegraphics[width=0.5\textheight]{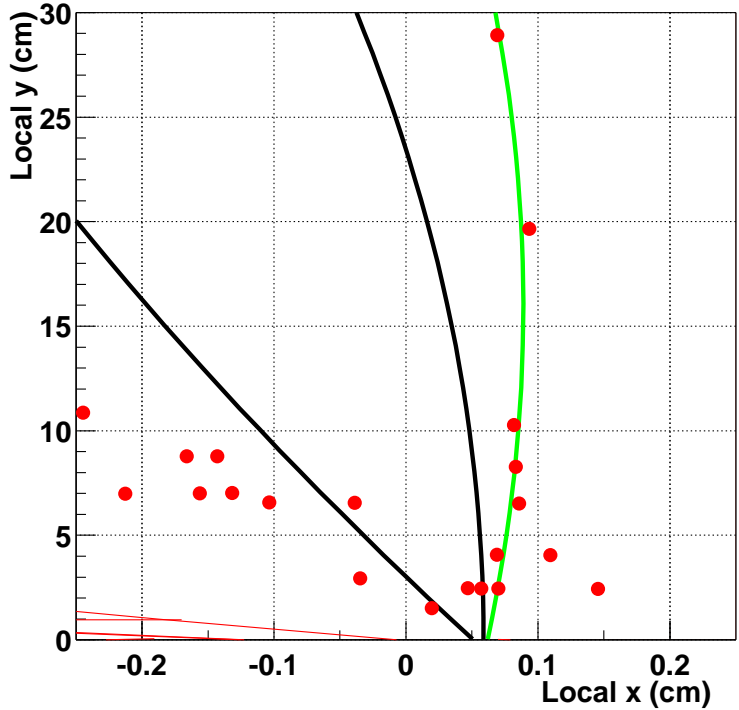}
\caption{An example of a lepton in the forward region from data where
one calorimeter seeded silicon track is
reconstructed. Shown is the xy-view of the silicon detector
with silicon hits (bullets). The pair of black tracks are the calorimeter seed
tracks projected into the silicon. 
The silicon pattern recognition algorithm is able to
clearly distinguish between the two charge hypotheses by attaching silicon
hits to the seed track on the right. The green track is the resulting silicon track. 
 \label{figtim}}
\end{figure}

These seeds are then projected into the silicon where hits are attached using 
the same pattern recognition algorithm which is used for COT seeded
tracks, see Fig.~\ref{figtim}.
Either of the seed tracks can result in a silicon track
if sufficient silicon hits are attached; a minimum of four hits are
required for this analysis.
If both seed tracks give rise to a silicon track,
the one with minimum $\chi^2/{\rm dof}$ is chosen.
The efficiency of this calorimeter-seeded algorithm is comparable to the
COT seeded algorithm.

Correct determination of the charge is important for the asymmetry
measurement. The charge mis-identification rate of the algorithm is
determined from the data using a sample of $Z\rightarrow e^+ e^-$ candidates.
One of the two leptons is required to be in the central region where its
charge is well identified by a matched COT track.
That determines the charge of the other lepton which is used to probe
the charge mis-identification rate as a function of $\eta$.

\begin{figure}
\includegraphics[width=0.6\textheight]{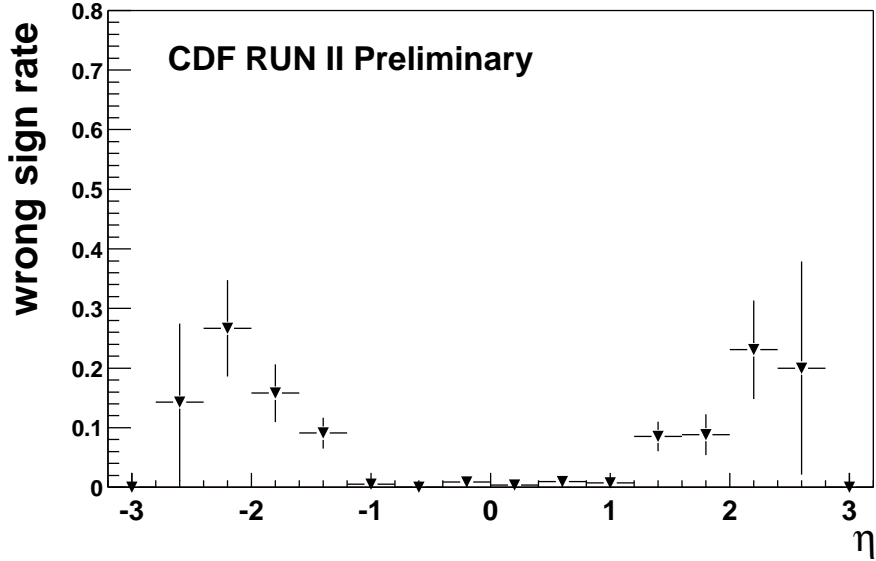}
\caption{Charge mis-identification rate measured with $Z \rightarrow e^+e^-$
data is plotted as a function of pseudo-rapidity.
\label{figcfr}}
\end{figure}

Fig. \ref{figcfr} shows the measured charge mis-identification rate
as a function of the lepton pseudo-rapidity.
In the central region the charge mis-identification is well below $2\%
$.
For $1.0 \le |\eta| \le 2.0$ it is between $10\%
$ and $15\%
$ rising up to $27\%
$ for $2.0<|\eta|\le2.6$. 
Improvements in the algorithm and the detector alignment
will reduce this rate to less than $10\%$ over the full $\eta$ range, 
but the performance is already sufficient for use in the asymmetry
measurement.

\section{Asymmetry Measurement} \label{secdatasets}

$W \rightarrow e\nu$ candidate events are selected from two online trigger
paths, one for the central region of the detector, $|\eta|<1$, and
another for the forward region.
The data sample was collected between February and September 2002 and
corresponds to an integrated luminosity of 32\,$\rm{pb}^{-1}$.
Selected events are required to satisfy the following criteria:
\begin{itemize}
\item exactly one electron candidate, passing the criteria described above,
with transverse energy $E_t > 15\,\rm{GeV}$
\item missing transverse energy $\not E_t > 30\,\rm{GeV}$
\item transverse mass, $50 \le M_t < 100\,\rm{GeV}$\footnote{The
transverse mass is given by
$M_t=\sqrt{[E_t(e)+E_t(\nu)]^2-[\vec{p}_t(e)+\vec{p}_t(\nu)]^2}$, where
$\vec{p}_t(e)$ and $\vec{p}_t(\nu)$ are the transverse momentum of the
electron and the neutrino.}.
\end{itemize}

\begin{figure}
\includegraphics[width=0.6\textheight]{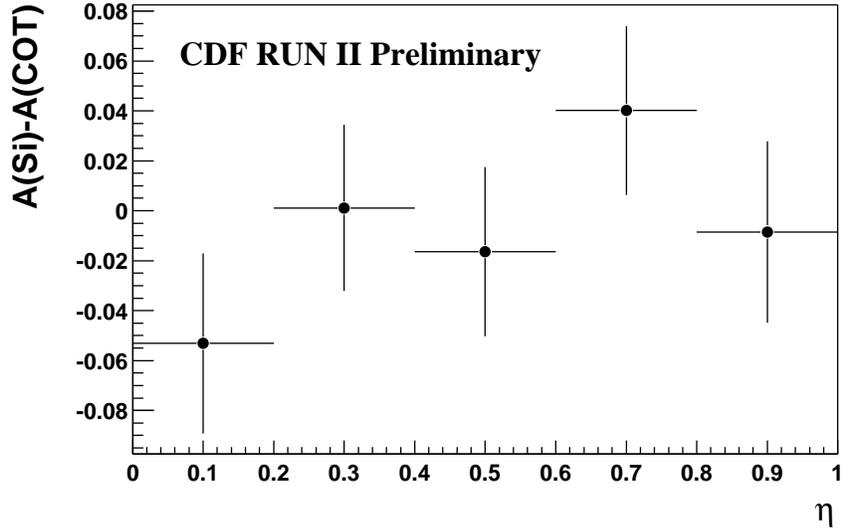}
\caption{Difference between the $W$ charge asymmetry determined with
calorimeter seeded silicon tracks and with COT tracks versus pseudo-rapidity. \label{figcotsi}}
\end{figure}

The calorimeter-seeded silicon tracking is used for the charge determination
in all $\eta$ regions. In the central region, $|\eta|<1$, the COT
tracking information is also available and is used as a cross-check.
Fig. \ref{figcotsi} shows the difference in that region between
the $W$ charge asymmetry measured with the silicon only tracking and with
the COT tracking. The two approaches yield consistent results.

\begin{figure}
\includegraphics[width=0.5\textheight]{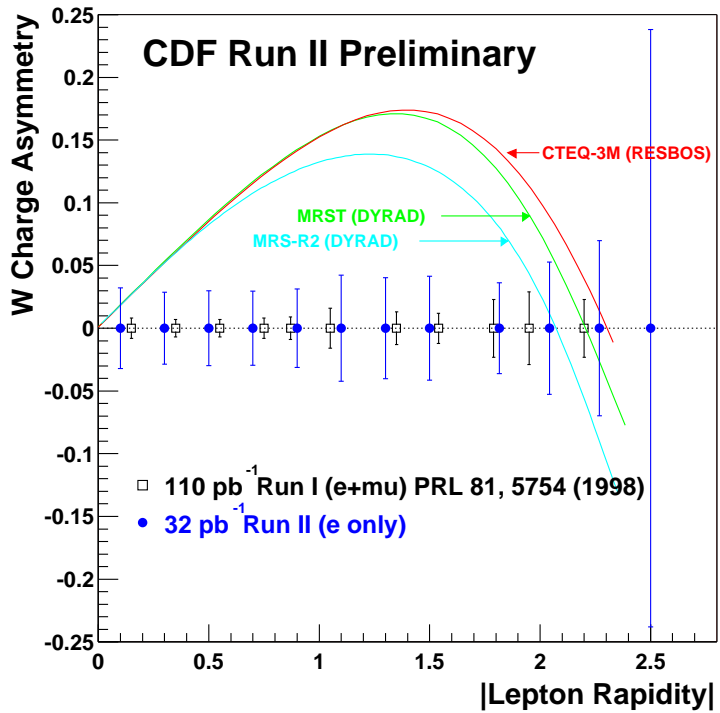}
\caption{$W$ charge asymmetry sensitivity with calorimeter-seeded
silicon tracks is compared between Run\,2 electrons (bullets)
and the Run\,1 measurement using both electrons and muons (open squares).
The central values of the data points are set to zero because the result
will not be unblinded until a larger data sample is collected and systematic
studies are completed.
The uncertainties shown include the statistical errors and
the systematic uncertainty from the charge mis-identification rate.
\label{figresults1}}
\end{figure}

Since the data sample collected so far is small, we do not yet consider
the actual measured asymmetry values. Rather, they are set to zero to
avoid potential bias and will not be unblinded until a larger data sample is
collected and systematic studies are completed.
Nonetheless, the sensitivity of the measurement can be determined.
This is shown in Fig. \ref{figresults1} where the
charge asymmetry is plotted as a function of $|\eta|$
but with the central values set to zero.
The uncertainties shown include the statistical errors and
the systematic uncertainty from the charge mis-identification rate.
For $|\eta| > 1.6$ the charge mis-identification uncertainty contributes
$50\%$ of the total error.
The figure compares this preliminary Run\,2 sensitivity to the
Run\,1 measurement ~\cite{Abe:1998} and to three parton density
functions, to indicate the range of variation in the parton
structure functions at high $\eta$.
 
\begin{figure}
\includegraphics[width=0.5\textheight]{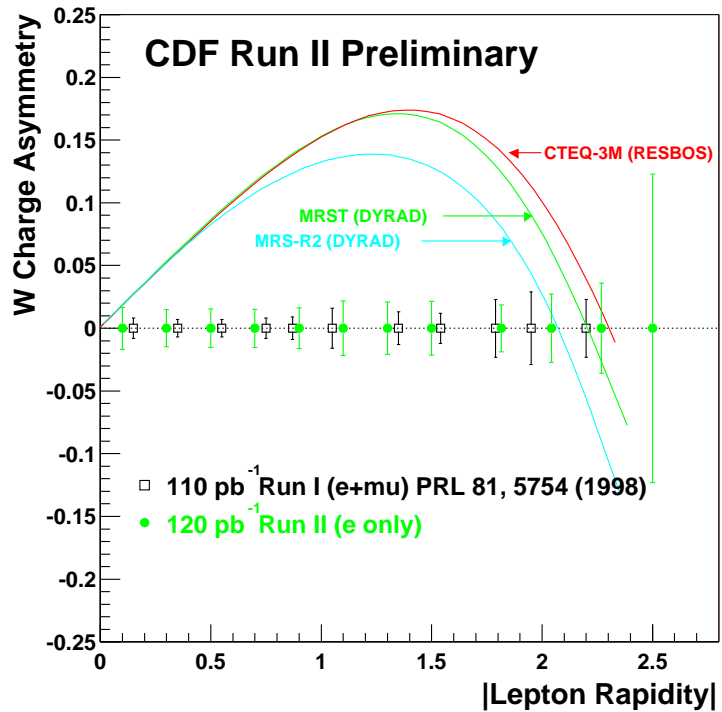}
\caption{A $120\,\rm{pb}^{-1}$ projection of the Run\,2 $W$ charge
asymmetry sensitivity (bullets) is compared to the Run\,1 measurement
(open squares) as a function of pseudo-rapidity.
\label{figresults2}}
\end{figure}

With only $32\,\rm{pb}^{-1}$ of Run\,2 data in the electron mode,
the sensitivity is not yet competitive with the existing Run\,1 measurement
obtained with both electrons and muons.
However, it can be used to provide a measure of the improvement in sensitivity
which will be obtained beyond just the ultimately higher integrated luminosity.
This is illustrated in Fig. \ref{figresults2} where the current uncertainties
are scaled to a $120\,\rm{pb}^{-1}$ sample which should be collected in the near
future. 
This projection is a conservative comparison in that it does not include the
effect of the muon mode from Run\,1 or the ongoing improvements in Run\,2
detector performance.
Nonetheless, it illustrates the gain in precision which will soon be obtained
in the $|\eta|>1$ region to further constrain the $u$/$d$ ratio
in the proton structure function. Furthermore the Run\,2 measurement
can extend the sensitivity up to up to $|\eta|=2.5$.

\begin{theacknowledgments}
I would like to thank the Division of Particles and Fields of the
Mexican Physical Society for the invitation to and the organization of
the X Mexican School of Particles and Fields. 
I thank Joel Goldstein, Joe Incandela, Tim Nelson, Rick Snider and
David Stuart for close collaboration and discussions. 
I thank the Fermilab staff and the technical staff of the participating 
institutions for their vital contributions. This work was
supported by the U.S. Department of Energy and National Science Foundation;
the Italian Istituto Nazionale di Fisica Nucleare; the Ministry of Education, 
Science and Culture of Japan; the Natural Sciences and Engineering Research
Council of Canada; the National Science Council of the Republic of China; 
and the A. P. Sloan Foundation.
\end{theacknowledgments}

% choose bibtex style depending on layout style and options used in
% sample:

%\doingARLO[\bibliographystyle{aipproc}]
%          {\ifthenelse{\equal{\AIPcitestyleselect}{num}}
%             {\bibliographystyle{arlonum}}
%             {\bibliographystyle{arlobib}}
%          }
%\bibliography{sample_cig}

\hyphenation{Post-Script Sprin-ger}

\end{document}